# Spotting 2-D Atomic Layers on Aluminum Nitride Thin Films

Hareesh Chandrasekar, Krishna Bharadwaj B, Kranthi Kumar Vaidyuala, Swathi Suran, Navakanta Bhat, Manoj Varma, and Srinivasan Raghavan*

*Centre for Nano Science and Technology, Indian Institute of Science, CV Raman Road, Bangalore, 560012, India*

Substrates for 2D materials are important for tailoring their fundamental properties and realizing device applications. Aluminum nitride films on silicon are promising large-area substrates for such devices in view of their high surface phonon energies and reasonably large dielectric constants. In this paper epitaxial layers of AlN on 2" Si wafers have been investigated as a necessary first step to realize devices from exfoliated or transferred atomic layers. Significant thickness dependent contrast enhancements are both predicted and observed for monolayers of graphene and $MoS_2$ on AlN films as compared to the conventional $SiO_2$ films on silicon, with calculated contrast values approaching 100% for graphene on AlN as compared to 8% for $SiO_2$ at normal incidences. Quantitative estimates of experimentally measured contrast using reflectance spectroscopy show very good agreement with calculated values. Transistors of monolayer graphene on AlN films are demonstrated, indicating the feasibility of complete device fabrication on the identified layers.

* sraghavan@cense.iisc.ernet.in

# I. INTRODUCTION

The right choice of substrates for 2D materials plays a major role in enabling their fundamental studies and device applications.[1-4] While $SiO_2$/Si is the most commonly used substrate for layered materials, diamond-like carbon,[5] silicon carbide[6] and hexagonal boron nitride[7] among others have all been employed for improving the performance of electronic devices based on few atomic layers. Remote interfacial phonon scattering remains a dominant scattering mechanism, under conventional device operating temperatures, leading to performance degradation in 2D layered transistors.[4, 8-10] Hence addressing its reduction imposes an important constraint in the choice of substrates for such devices.[5, 7] Aluminum nitride (AlN), a member of the group III-A nitride family (along with GaN and InN) is a highly stable wide band gap semiconductor. AlN is an attractive substrate for atomic layered devices due to its high energy of surface optical phonon modes which reduces remote interfacial phonon scattering, giving rise to improved performance in such devices.[11] It has also been very recently predicted that AlN and BN substrates offer the optimal trade-off between surface optical phonon scattering and the impurity scattering limited mobility for monolayers of $MoS_2$ at room temperature thus providing high mobilities and a reasonably high dielectric constant, both at the high carrier densities necessary for device operation.[12] The high thermal conductivity of AlN would also enable efficient heat dissipation thus restricting self-heating effects.[10, 13, 14] AlN films can be integrated reliably[15] on large-area (6" and 8") Si wafers[16, 17] making them technologically and commercially significant for devices of layered materials. Additionally, the possibility of realizing 2D hexagonal AlN itself has been investigated by first principles DFT calculations[18], which if experimentally realized offers intriguing possibilities for extending the family of 2D materials and heterostructures.

$SiO_2$ films on Si remain the standard substrate for layered materials primarily due to the ability to spot even monolayers atop such films under an optical microscope.[19-24] Therefore, fabricating devices on 2D materials placed on AlN films as substrates would depend to a very large extent on their facile identification post exfoliation or transfer. Optical microscopy based quantification of the thickness of layered materials has also been demonstrated by means of techniques such as contrast spectra[25] and color differences[21] to name but two. Reliable thickness determination of graphene layers using a selection of such methods under a unified theoretical framework has also been summarized by Ouyang et al.[26], and can be employed to determine the number of layers of arbitrary 2D materials on any substrate accurately. It is noteworthy that irrespective of the actual method employed for thickness dependent optical identification, monolayers of layered materials are always the hardest to identify on any substrates due to their lower contrast.

In this Letter, we report the optical identification of monolayers of graphene and MoS2 on AlN thin films on Si. Approaches to enhance optical contrast typically involve stacking multiple layers of materials having distinct refractive indices to achieve destructive interference from the various interfaces.[27, 28] Such an anti-reflection effect can also be achieved using a single AlN _lm on Si. We show that the theoretical Michelson contrasts of graphene and MoS2 on AlN approach their maximum possible values (±100%) at normal incidence, due to the anti-reflection effect of AlN/Si. Experimentally, even at non-normal incidences as in a simple optical microscope, graphene and MoS2 monolayers on AlN/Si show significant enhancement compared to SiO2/Si, which would enable their facile identification on these substrates. We perform optical reflectance spectroscopy to compare the experimentally observed contrast with theoretical predictions and obtain good agreement.

## II. EXPERIMENTAL DETAILS
### A. Growth, transfer and characterization of AlN, graphene and $MoS_2$

AlN films of 50 nm, 100 nm, 150 nm and 200 nm (targeted thicknesses) were deposited on Si (111) using Metal Organic Chemical Vapor Deposition (MOCVD) at 1050°C using tri-methyl aluminum and ammonia as precursors, details reported elsewhere.[15] While the reason for choosing these particular thickness values will be clarified below, we need to mention that the lower limit of film thickness is only determined by the ability to obtain continuous layers of AlN on silicon, which typically grows in the Volmer-Weber (3D) growth mode. Ellipsometry measurements were performed using a J.A Woolam Co. M2000U to determine the optical constants and film thicknesses of these four AlN samples which were found to be 54 nm, 97 nm, 159 nm and 202 nm respectively, with upto a 10% thickness variation across the entire silicon wafer. Graphene and $MoS_2$ growth were carried out in a home-built CVD reactor. Graphene was grown on copper foils at 1000°C and $MoS_2$ on $SiO_2$ substrates at 850°C using methane, and molybdenum hexacarbonyl and hydrogen sulphide as precursors respectively. A PMMA transfer process was employed to transfer both graphene and $MoS_2$ onto AlN followed by acetone treatment to remove PMMA residues. Raman measurements were done in a Horiba LabRAM HR with a 532 nm laser.

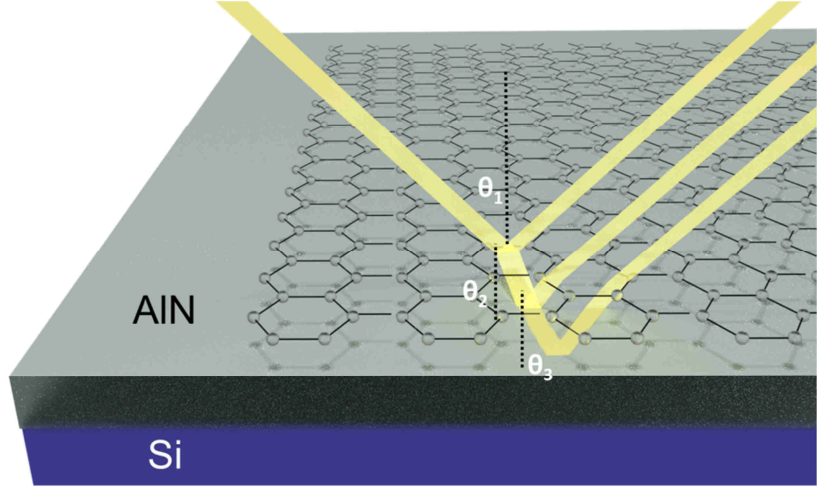

Figure 1. Schematic of light reflection and refraction at the three interfaces - air/2D, 2D/AlN and AlN/Si. $\theta_1$, $\theta_2$ and $\theta_3$ refer to the angle of incident light into each of the three layers respectively.

**B. Optical Contrast and Reflectance measurements**

Optical contrast was calculated using the multilayer reflectance model by calculating reflectivity and hence the reflectance from the 2D-dielectric-Si and dielectric-Si cases under normal incidence.[20] Optical constants of graphene and $MoS_2$ as measured by spectroscopic ellipsometry, reported in literature[29, 30] were used along with thickness values of 0.34 nm and 0.67 nm respectively. The contrast was also calculated assuming a Gaussian distribution of reflected intensity over the beam angle corresponding to a numerical aperture of 0.8 (50x objective). Samples were imaged, using a white light source, in areas completely covered with 2D layers and on the bare substrate, under an Olympus BX51M upright microscope. All observations were carried out in the bright field mode without any filters. For spectroscopic measurements, the optical signal was collected with a fiber focused through the microscope eyepiece and directed to an Ocean Optics spectrometer HL2000. Optical images were analyzed using the open source ImageJ processing software[31] and pixel intensities of the red, green and blue color channels were extracted using the Split Channels function.

**III. RESULTS AND DISCUSSION**

We define the Michelson contrast as

$C_M = (I_S - I_{2D})/(I_S + I_{2D})$ (1)

where $I_S$ refers to reflected light intensity from the bare substrate and $I_{2D}$ the reflected intensity from the 2D layer, with intensities calculated as indicated by Jung et al.[20] The reflected intensities are given by $I_{2D} = r_{2D} r_{2D}*$ with $r_{2D}$ being the amplitude of reflected light and similarly for $I_S$. The reflected amplitudes are in turn calculated as

$$r_{2D} = \frac{r_1 + r_2 \exp(-2i\delta_2) + r_3 \exp(-2i(\delta_2 + \delta_3)) + r_1 r_2 r_3 \exp(-2i\delta_3)}{1 + r_1 r_2 \exp(-2i\delta_2) + r_2 r_3 \exp(-2i\delta_3) + r_1 r_3 \exp(-2i(\delta_2 + \delta_3))} \quad (2)$$

$r_1$, $r_2$ and $r_3$ are the average reflected amplitudes for s- and p- polarized light from the air/2D, 2D/AlN and AlN/Si interfaces and the phase difference due to the m$^{th}$ layer of thickness $t_m$, real and imaginary components of refractive index being $n_m$ and $k_m$ respectively and $\delta_m$, the angle of incident light into the layer is given by $\delta_m = 2\pi t_m(n_m - ik_m) \cos\theta_m /\lambda$ as seen from Fig.1. Similarly for $I_S = r_S r_S*$ with the reflected amplitude from the substrate given by

$$r_S = \frac{\hat{r}_1 + \hat{r}_2 \exp(-2i\hat{\delta}_2)}{1 + r_1 r_2 \exp(-2i\hat{\delta}_2)} \quad (3)$$

and $\hat{r}_1$ and $\hat{r}_2$ are the average reflected amplitudes from the air/AlN and AlN/Si interfaces in the absence of the 2D layer and $\hat{\delta}_2$ the phase difference is denoted as in Eq. (2).

Positive values of $C_M$ refer to absorptive contrast, i.e. the material appears darker than the substrate; negative values indicate reflective contrast, i.e. material looks brighter against the substrate. We see that maximum values of contrast, $C_M = +1$ and -1 are obtained when the 2D/dielectric/substrate or the dielectric/substrate combination satisfies the condition for complete destructive interference.

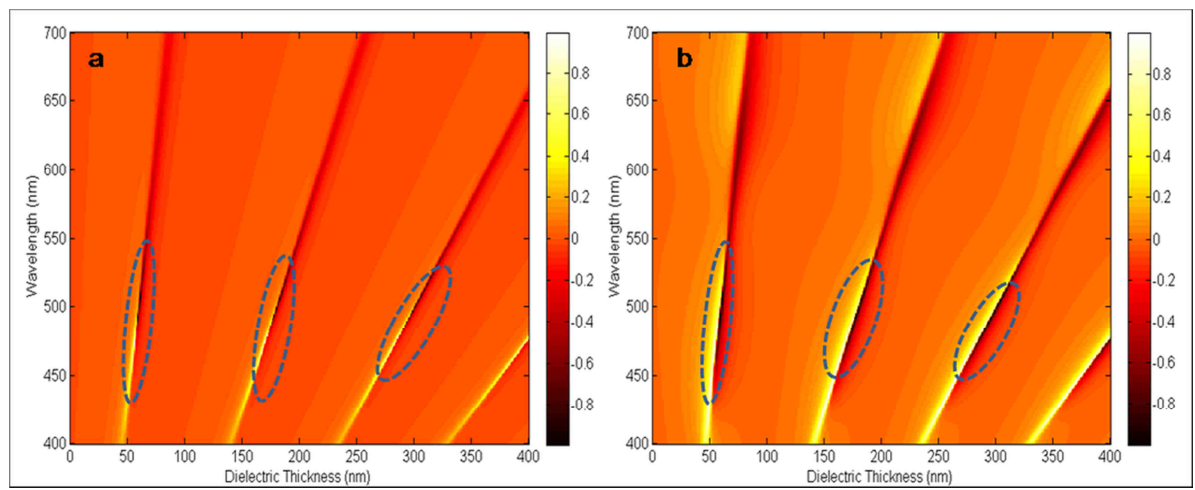

Figure 2. Contrast maps for monolayers of a) graphene and b) MoS$_2$ at different wavelengths and AlN film thickness under normal incidence. Maximum possible contrast values of ±1 can be discerned due to the anti-reflection nature of these films. Dotted lines indicate high contrast regions for up to 10 nm thickness bands making them experimentally accessible as discussed in the text.

While this condition never occurs in the visible spectrum for the range of refractive indices of $SiO_2$ (1.47-1.45), the higher refractive index of AlN (2.19-2.14) enables such an anti-reflection condition for the appropriate wavelength-thickness combinations during normal incidence.

Figs. 2a and 2b show the calculated contrast maps for graphene and $MoS_2$ in the visible spectrum, under normal incidence, for a range of AlN thicknesses. Contrasts of up to ±1, the maximum possible value, can be obtained for AlN/Si corresponding to a 12x and 4x contrast enhancement over $SiO_2$/Si at normal incidence due to the anti- reflection nature of both the 2D/AlN/Si ($I_{2D}$= 0, CM= +1) and AlN/Si ($I_S$= 0, CM= -1) interfaces at particular wavelength-thickness combinations. We also observe that contrast values much greater than those for $SiO_2$ can be perceived for a wide range of wavelengths (~400-470 nm) and for as much as 10 nm bands of thicknesses (for example, 47-57 nm at 450 nm wavelength) which enables their realization, using appropriate filters at near normal incidence, with relatively straightforward modifications of the standard microscope optical configuration. However as we show, even without any modification of the microscope, i.e. with white light illumination and a standard objective, the contrast observed is enhanced with respect to $SiO_2$ films. Eqs. (2) and (3) can also be extended to estimate the contrast for multi-layers of these materials as well. The contrast for bi- and tri-layers of graphene and $MoS_2$ on AlN is also observed to be similar to those for monolayers, but with the regions exhibiting higher contrast now spanning a much wider range of wavelength-thickness as expected for thicker samples (see Figs. S1 and S2 in supplementary material).

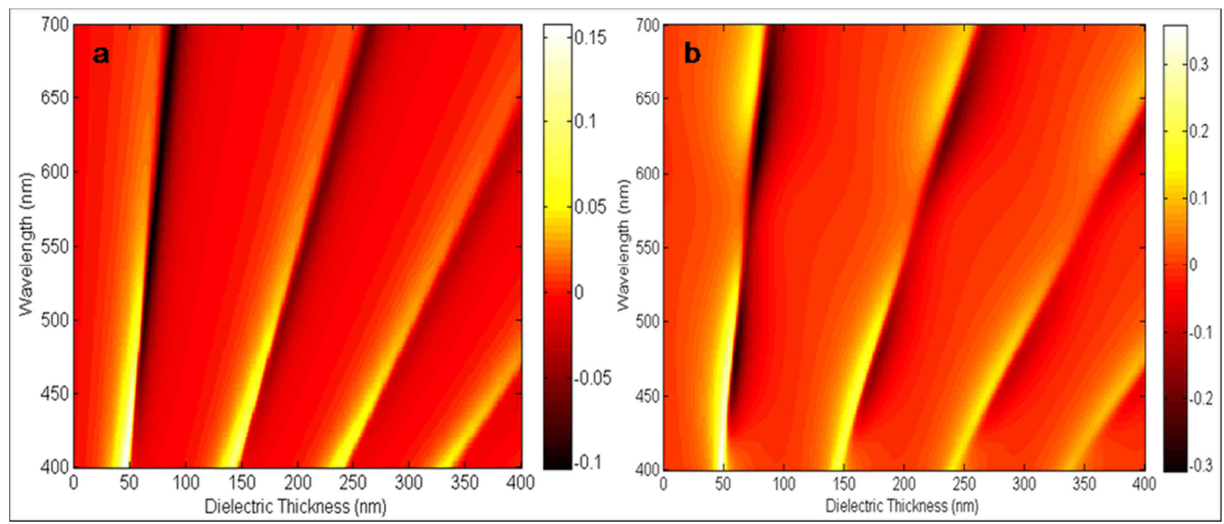

Figure 3. Contrast maps for monolayer a) graphene and b) MoS$_2$ at different wavelengths and AlN film thicknesses for an experimentally pertinent 50x objective with a NA=0.8. Both positive and negative contrast can be discerned unlike for SiO$_2$/Si where the contrast is predominantly absorptive.

Visualization of 2D materials is typically done using microscope objectives resulting in a decrease of optical contrast due to the numerical aperture of the objective. Figs. 3a and 3b show the contrast maps of Fig. 2 taking into account the numerical aperture (NA=0.8) of the 50x objective, corresponding to our experimental setup. The maximum contrast attainable in this case corresponds to a 2x and 1.4x enhancement over SiO$_2$/Si under similar imaging conditions (see Fig. S3 and S4 in supplementary material[32] for contrast maps on SiO$_2$/Si). Unlike the case of SiO$_2$/Si where the contrast is predominantly absorptive, we distinguish regions of both reflective and absorptive contrast for graphene and MoS$_2$ on AlN. That is, 2D layers look darker (<50 nm of AlN), brighter (50-100 nm AlN) and for some AlN thicknesses (140-200 nm) can transition from appearing brighter to darker against the substrate, depending on the wavelength of interest. Different AlN thicknesses were used for contrast studies in order investigate these effects and we show clear evidence for all three of these features below.

Figs. 4 and 5 show optical images of graphene and MoS$_2$ for four different AlN thicknesses - 54 nm, 97 nm, 159 nm and 202 nm. Their corresponding intensity profiles on red, green and blue channels along the line A-B shown and Raman spectra at points A and B, on the substrate and 2D layer respectively, are also included. For AlN thicknesses of 54 nm and 159 nm, the 2D layer appears darker against the substrate (Figs. 4a and 4c; 5a and 5c), similar to the conventional SiO$_2$/Si case. This is also indicated by the pixel intensities shown alongside (Figs. 4e and 4g; 5e and 5g) confirming the trends in the contrast maps of Fig. 3. In contrast, the 2D layer appears brighter than the substrate for 97 nm films (Figs. 4b and 5b). This is also seen from the plot of pixel intensities (Figs. 4f and 5f), in agreement with the contrast maps in Fig. 3.

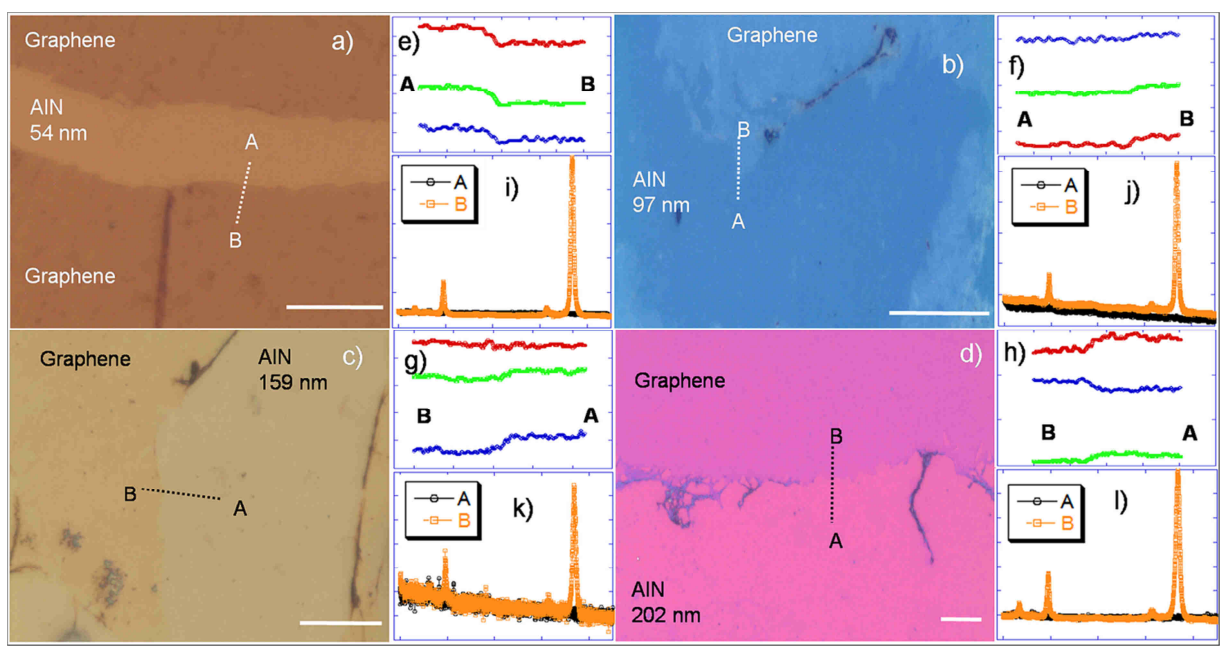

Figure 4. (a-d) are optical images of graphene on AlN films of 54, 97, 159 and 202 nm thicknesses respectively, scale bars correspond to 20 mm. (e-h) show their intensity profiles (in arbitrary units) on the red, green and blue channels along the line A-B, point A being on the substrate and B on graphene. (i-l) are corresponding Raman spectra of the points A and B, the x-axis is wave number in cm$^{-1}$ from 1200 to 3000, y-axis is intensity in arbitrary units. The G peak is observed at ~1580 cm$^{-1}$ and a more intense 2D peak at ~2650 cm$^{-1}$ indicating the presence of monolayer graphene.

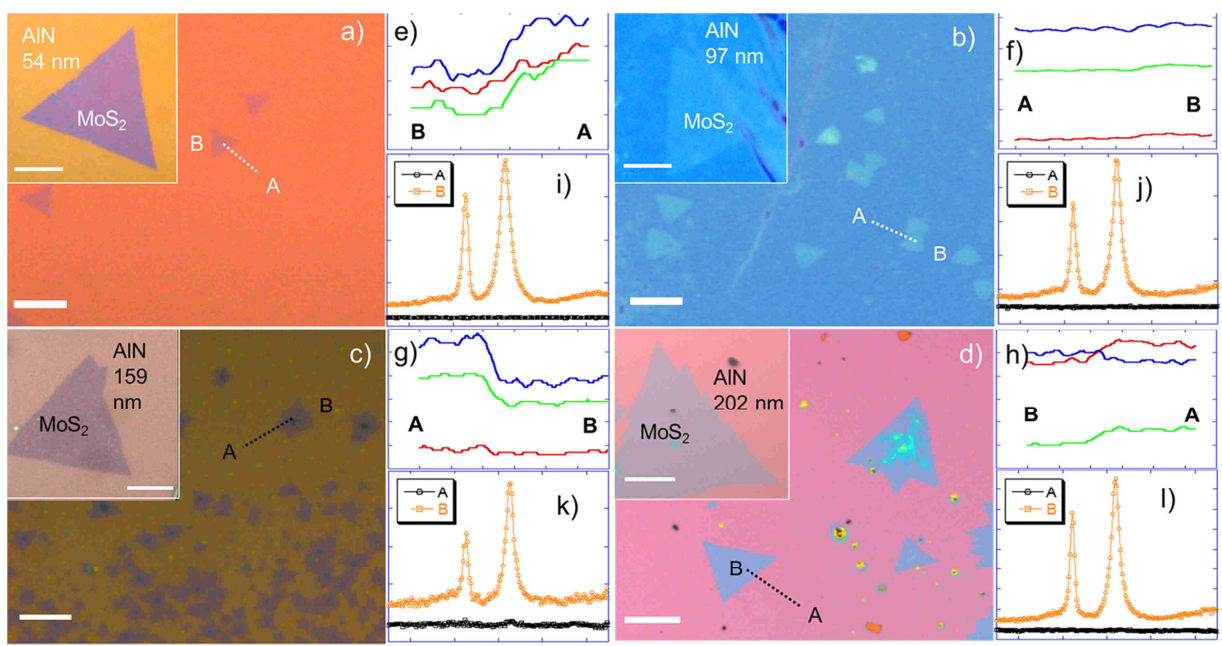

Figure 5. (a-d) are optical images of MoS$_2$ flakes on AlN films of 54, 97, 159 and 202 nm thicknesses respectively; inset shows a triangular domain of monolayer MoS$_2$ on AlN. Scale bars correspond to 20 mm (5 mm in the inset). (e-h) show their intensity profiles (in arbitrary units) on the red, green and blue channels along the line A-B, point A being on the substrate and B on MoS$_2$. (i-l) are corresponding Raman spectra on the points A and B, indicating clearly the absence and presence of monolayer MoS$_2$. The x-axis ranges from 350 to 450 cm$^{-1}$

and the y-axis is intensity in arbitrary units. The $E_{2g}^1$ peak is observed at ~384 cm$^{-1}$ and the $A_{1g}$ peak at ~403 cm$^{-1}$ indicating the presence of monolayer MoS$_2$.

For an AlN thickness of 202 nm, the contrast should change from reflective to absorptive with increase in wavelength, i.e. the 2D layer looks brighter against the substrate in the blue range of wavelengths and darker for wavelengths > 550 nm (see Fig. 3). Such a feature is clearly evident in the pixel intensities (Figs. 4h and 5h), with the blue channel displaying higher intensities on the 2D layer (point B), whereas the red and green channels show higher pixel intensities on the substrate (point A). Raman spectra as evidenced by the 2D and G peaks shown in Figs. 4i-4l clearly indicate the presence of monolayer graphene (point B) on all samples and the lack thereof on the substrate (point A).[33] Similarly, Raman spectra in Figs. 5i-5l confirms the presence of single-layered MoS$_2$ on all samples as evidenced by <20 cm$^{-1}$ difference between the $E_{2g}^1$ and $A_{1g}$ modes.[34]

In order to quantify and compare these qualitative trends in the observed contrast with the predicted values, we performed spectroscopic reflectance measurements on large area, contiguous single layers of graphene and MoS$_2$. The reflectance curves measured from AlN/Si and the 2D/AlN/Si regions clearly indicate their anti-reflection nature by approaching zero reflectance for the predicted thickness-wavelengths as shown in Figs. S5 and S6.[32] Figs. 6 and 7 display the contrast extracted from reflectance in the 450-700 nm wavelength range for graphene and MoS2 respectively at the four AlN thicknesses discussed above. The theoretical contrast band, accounting for film thickness variations, is plotted alongside for comparison. We see that the two curves show good agreement over the entire wavelength range of measurement for all AlN thicknesses in this study. The reflectance measurement consistently overestimates the contrast as compared to the predicted values. This can be attributed to the presence of wrinkles and folds in the 2D material, and mild PMMA residues from the transfer process, commonly observed during the transfer of such large area layers (~5 mm x 5 mm in our case).

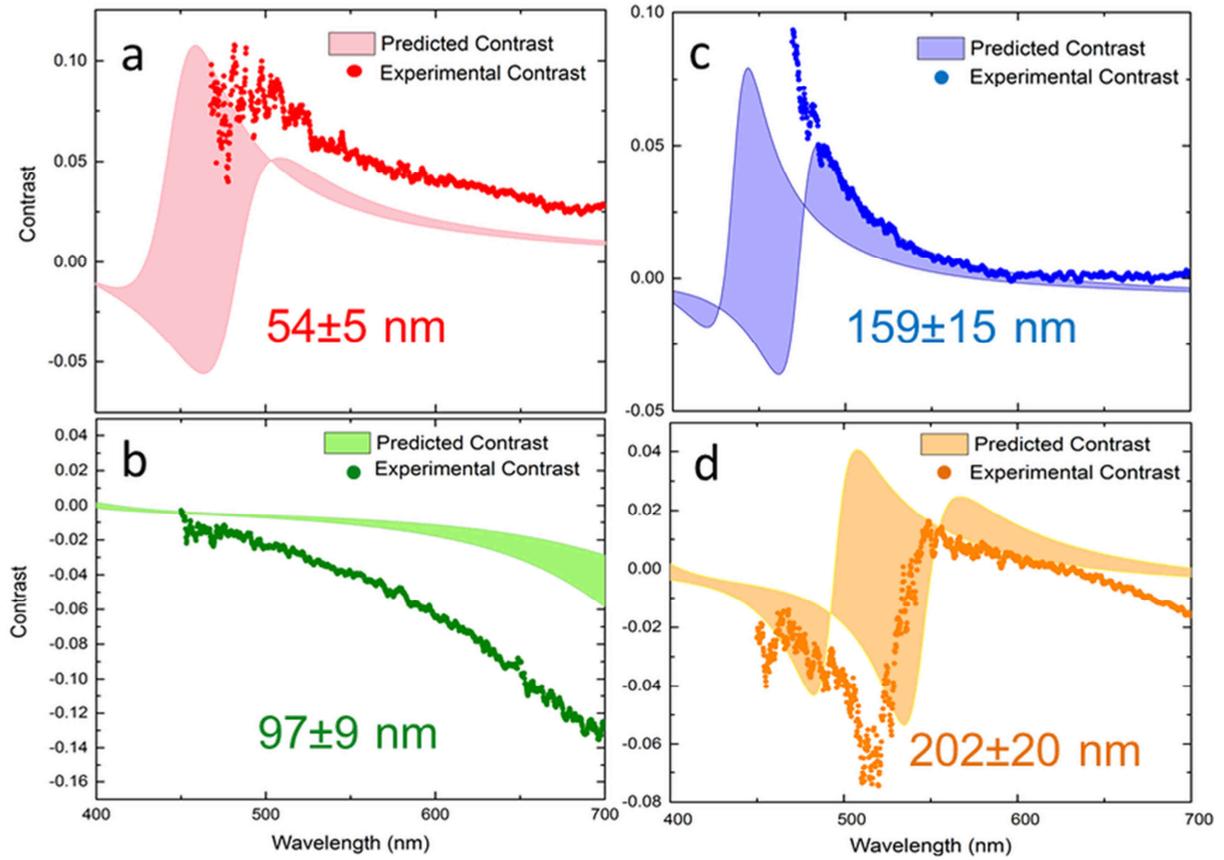

Figure 6. Experimental contrast values extracted from reflectance spectroscopy as compared to theoretically predicted values for graphene on AlN films of thicknesses of a) 54 nm, b) 97 nm, c) 159 nm and d) 202 nm. The predicted values have been extracted from the contrast maps shown in Fig. 2 for the appropriate thicknesses. The bands in the theoretical contrast correspond to thickness variations in the AlN films as indicated in each figure.

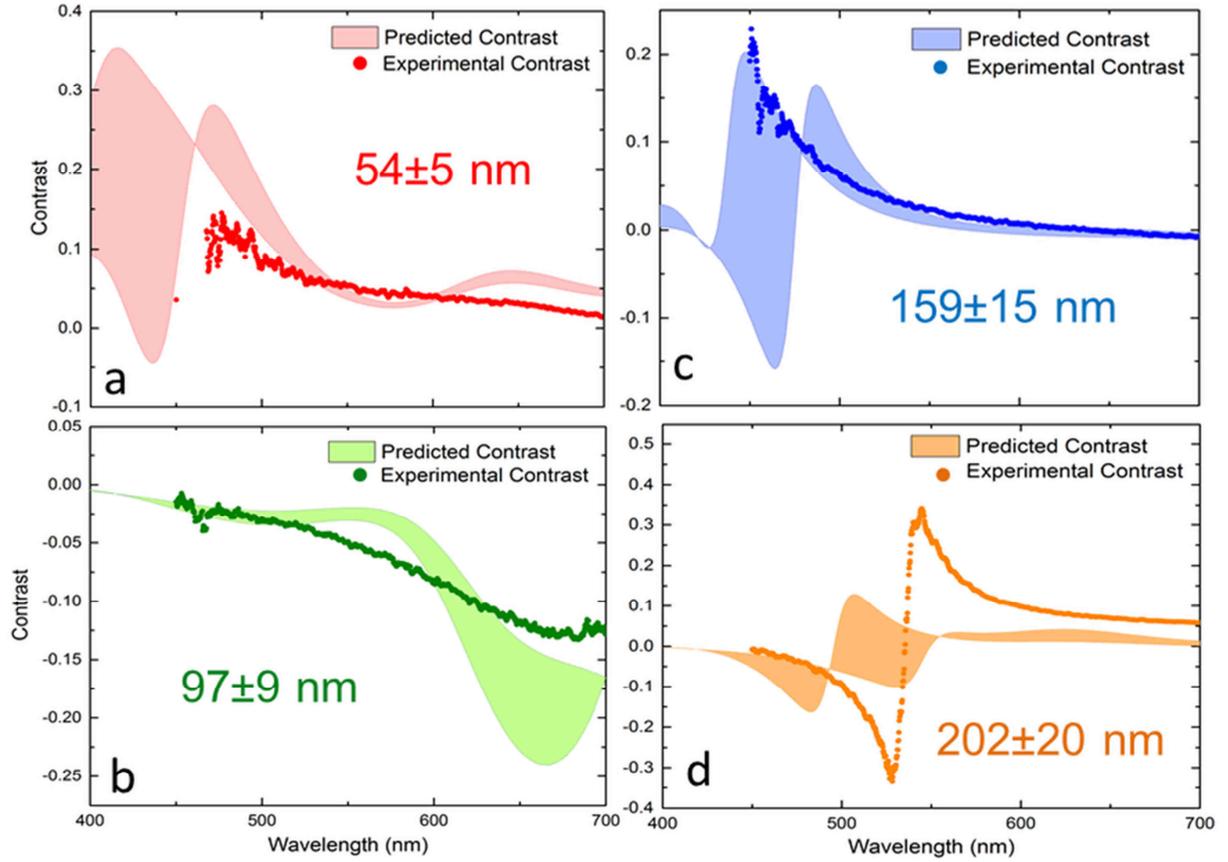

Figure 7. Experimental contrast values extracted from reflectance spectroscopy as compared to the theoretically predicted values for $MoS_2$ on AlN films of thicknesses of a) 54 nm, b) 97 nm, c) 159 nm and d) 202 nm. The predicted values have been extracted from the contrast maps shown in Fig. 2 for the appropriate thicknesses. The bands in the theoretical contrast correspond to thickness variations in the AlN films as indicated in each figure.

In order to demonstrate the feasibility of complete device fabrication on the 2D layers thus spotted, lithographically patterned graphene strips of 10 mm x 80 mm on a 202 nm AlN film were etched using oxygen plasma in an RIE, followed by Cr/Au contact deposition using e-beam evaporation. Fig. 8 shows an optical micrograph of the transistor along with its corresponding $I_d$-$V_g$ curves at room temperature, indicating the standard device fabrication flow for $SiO_2$ can be employed as is for the 2D layers thus spotted on AlN/Si. In order to further demonstrate the advantage of AlN films over $SiO_2$, the channel mobilities for graphene transistors shown in Fig. 8 have been extracted and compared to reference devices, of identical dimensions subject to the same fabrication procedure described above, on 300 nm $SiO_2$ films. A constant mobility model[35] has been used to fit the transfer characteristics using a single channel mobility for electrons and holes as shown below.

$$R_{tot} = R_{contact} + \frac{L/W}{\sqrt{q^2 n_0^2 + C_{dielec}^2 (V_G - V_{Dirac})^2}\,\mu} \qquad (4)$$

where $R_{tot}$ is the total measured resistance of the graphene transistor, $R_{contact}$ is the contact resistance, L and W are the length and width of the graphene channel, $n_0$ is the carrier concentration at the Dirac point, $C_{dielec}$ is the areal capacitance of the dielectric, $V_G$ is the applied gate bias, $V_{Dirac}$ is the Dirac point voltage and µ is the carrier mobility and $R_{contact}$, $n_0$ and µ are the fit parameters.

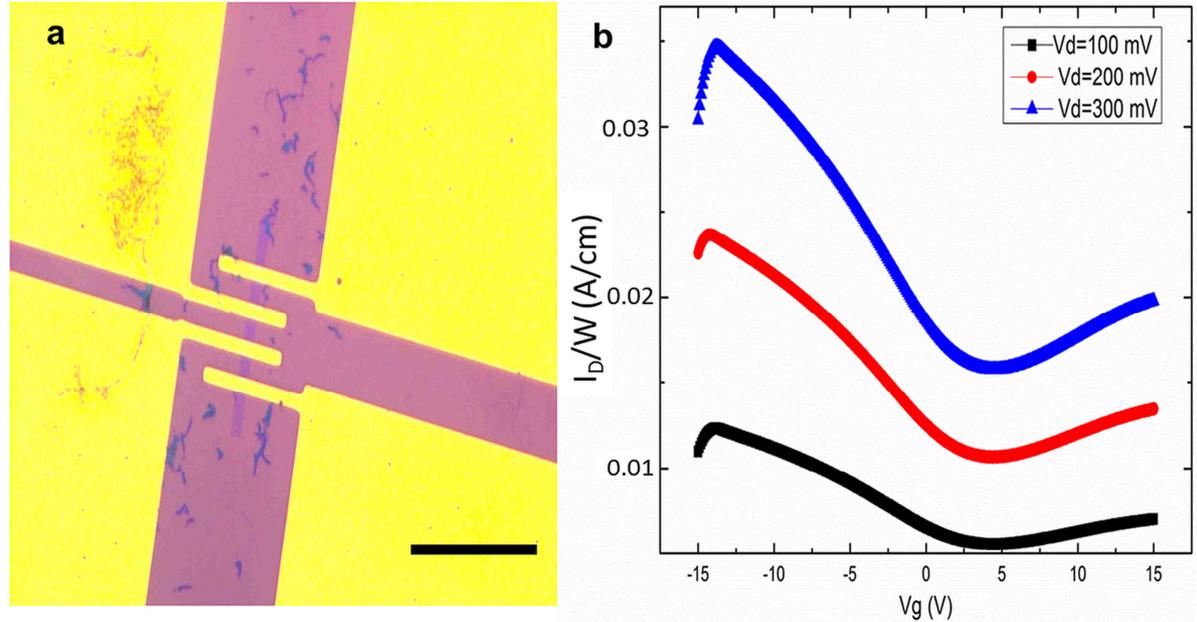

Figure 8. a) Optical image of a monolayer graphene transistor (10 mm x 80 mm) on 202 nm thick AlN deposited on a highly doped p-type Si as back gate. The rectangular graphene channel can be clearly seen against the substrate. The scale bar indicates 50 mm. b) Transfer characteristics ($I_d$-$V_g$), at room temperature, in the back gated configuration for drain voltages of 100 mV, 200 mV and 300 mV.

The channel mobilities in case of graphene devices on AlN are extracted to be 18240±1216 $cm^2$/Vs as compared to 8431±520 $cm^2$/Vs on $SiO_2$. This corresponds to a greater than 2-fold increase in channel mobility, at room temperatures, over $SiO_2$ films which can be attributed to the higher surface phonon energy of AlN and clearly indicates the advantage of using AlN films as substrates for devices of 2D layers.

## IV. CONCLUSIONS

In conclusion, we have shown that 2D materials - graphene and $MoS_2$ - can be optically identified on AlN films on Si with enhanced contrast as compared to $SiO_2$/Si, with ~50 nm of AlN offering the best possible contrast across the entire range of optical wavelengths. That the predicted contrasts can approach their maximum possible values of ±1 is due to the anti-reflection nature of these AlN/Si films. Quantitative estimation of contrast using reflectance spectroscopy shows good agreement with the predicted values at various AlN thicknesses for the entire visible spectrum, confirming the better visibility of 2D layers on AlN films. We have also realized single-layer graphene transistors on AlN/Si demonstrating the viability of further device processing on the 2D layers thus spotted. We believe such facile identification allows further investigations of atomic layers on AlN, especially in view of its promise as a substrate for device applications.

ACKNOWLEDGMENTS

The authors would like to acknowledge financial support from the Department of Science and Technology, Government of India under Grant No. SR/S2/CMP-02/2007 and the TUE project for the development of Nanoscience and Technology. The authors would like to acknowledge the Micro and Nano Characterization Facility at the Centre for Nano Science and Engineering, Indian Institute of Science for access to Raman spectroscopy and microscopy facilities.

## Supplementary Material

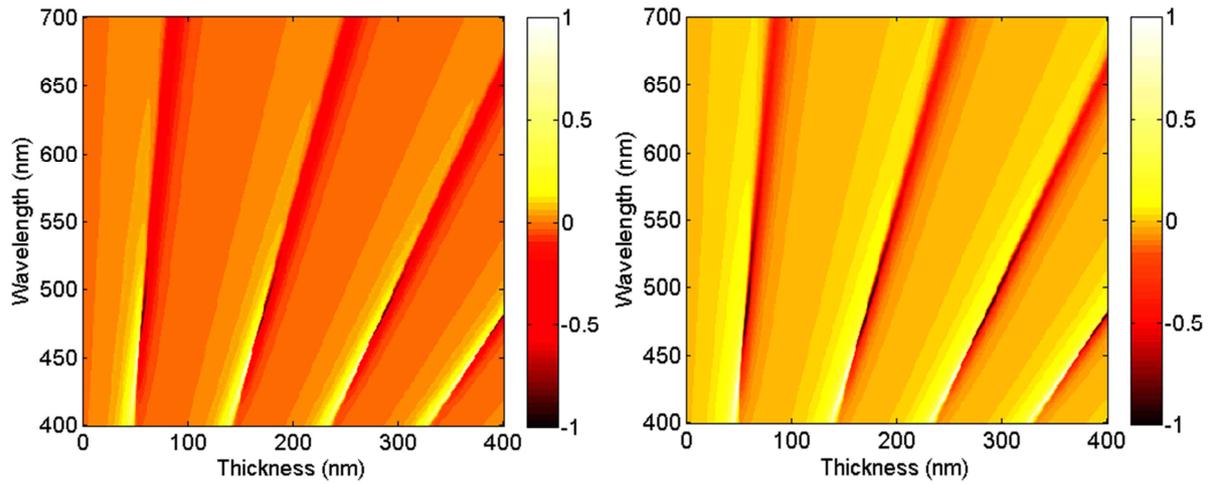

**Fig. S1** Contrast maps for bi-layer (left) and tri-layer (right) graphene for various AlN thickness-wavelength combinations.

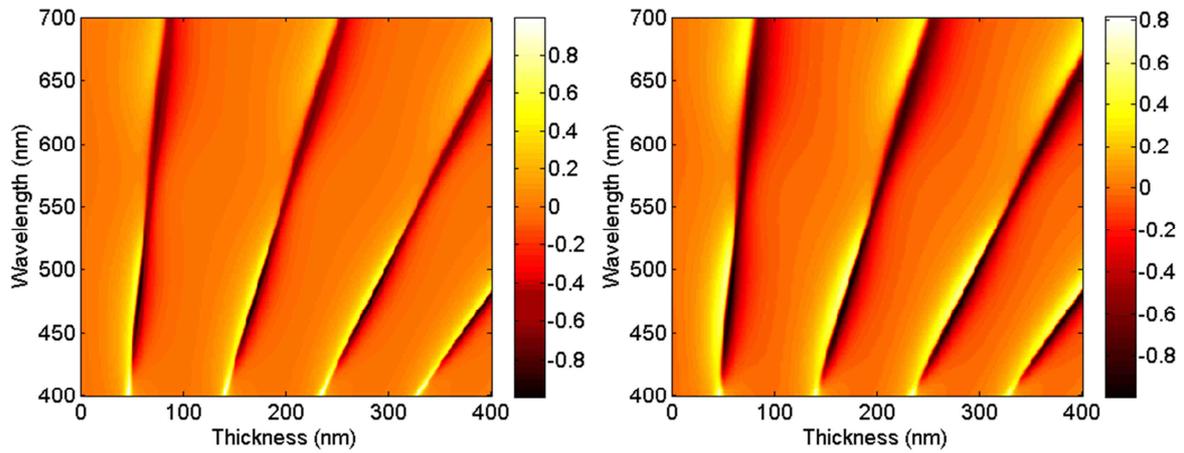

**Fig. S2** Contrast maps for bi-layers (left) and tri-layers (right) of $MoS_2$ for various AlN thickness-wavelength combinations.

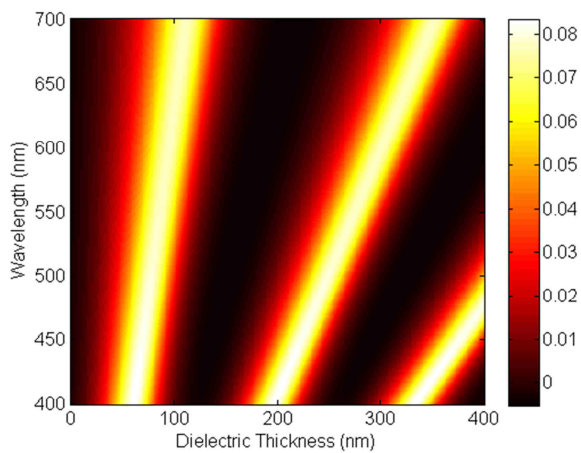

**Fig. S3** Contrast map for monolayer graphene on $SiO_2$/Si across the visible spectrum for different film thicknesses, accounting for a numerical aperture of 0.8.

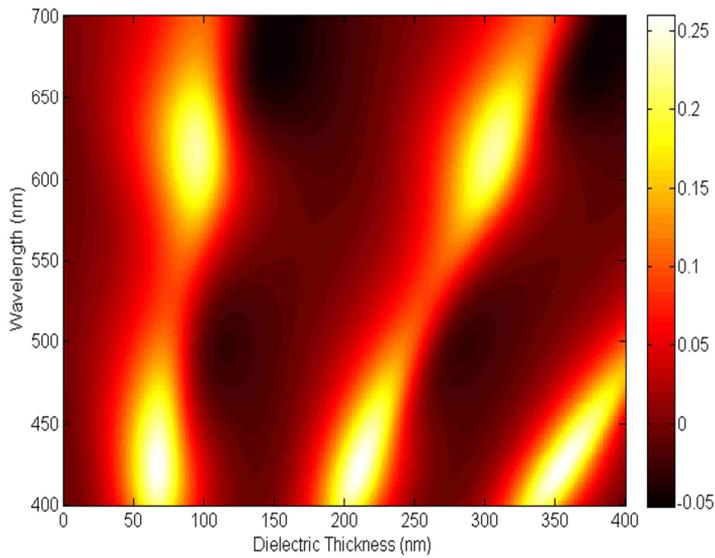

**Fig. S4** Contrast map for monolayer MoS$_2$ on SiO$_2$/Si across the visible spectrum for different film thicknesses, accounting for a numerical aperture of 0.8.

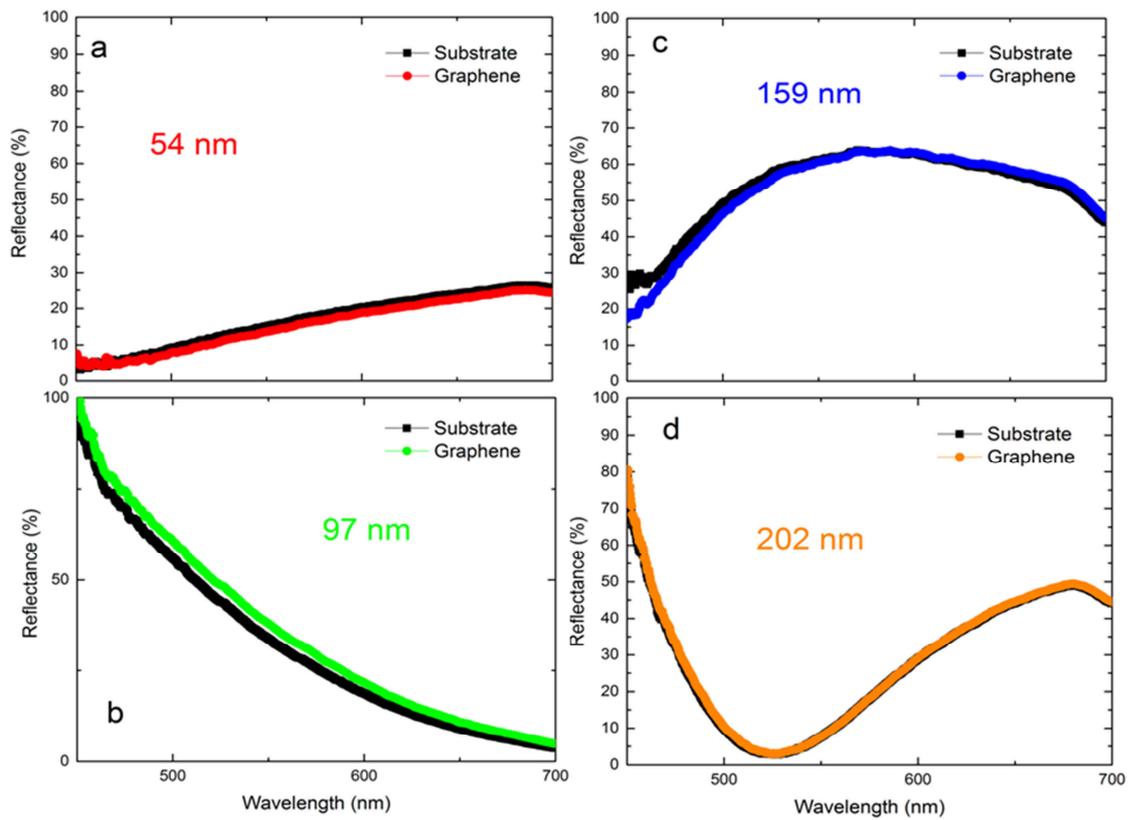

**Fig. S5** Reflectance versus wavelength as measured on graphene and bare AlN for different AlN thicknesses indicated in each graph, a) 54 nm, b) 97 nm, c) 159 nm and d) 202 nm. The onset of the anti-reflection behavior is clearly visible in the 202 nm case around 520 nm.

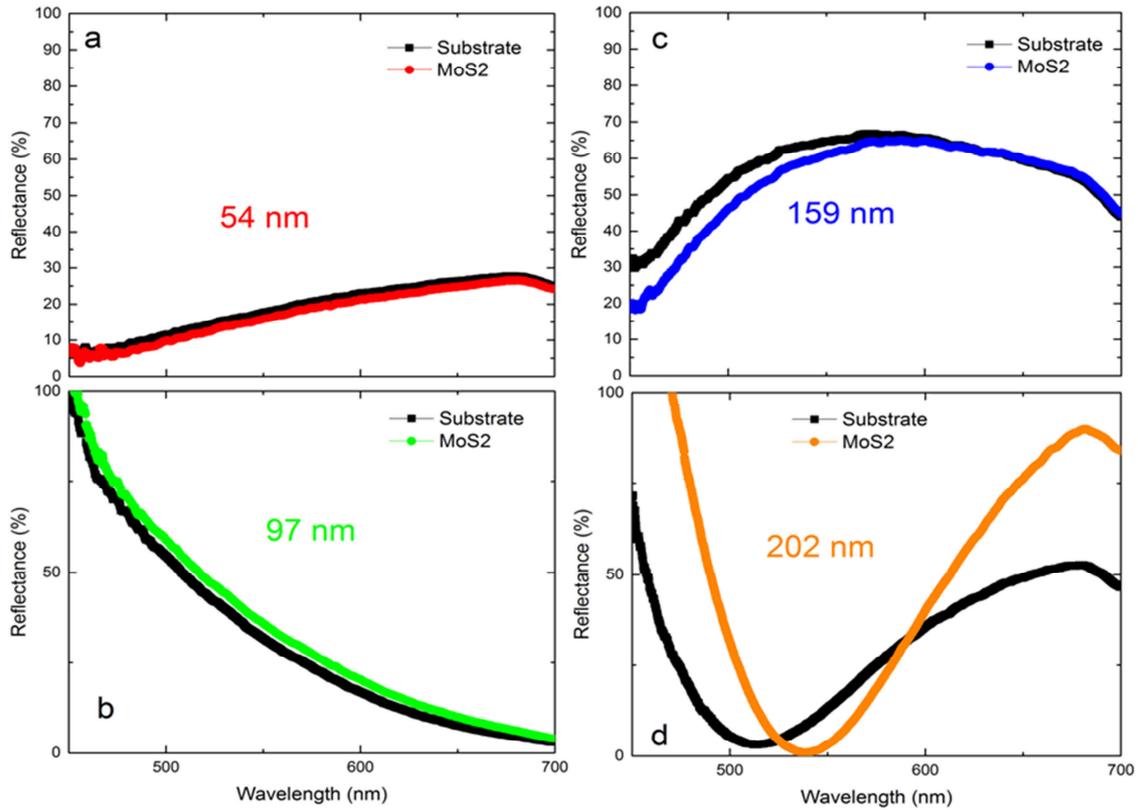

**Fig. S6** Reflectance versus wavelength as measured on MoS$_2$ and bare AlN for different AlN thicknesses indicated in each graph, a) 54 nm, b) 97 nm, c) 159 nm and d) 202 nm. The onset of the anti-reflection behavior is clearly visible in the 202 nm case at 520 nm for bare AlN which is then shifted by ~20 nm to 540 nm upon adding a monolayer of MoS$_2$.